\documentclass[aps]{revtex4}
\usepackage{graphicx}
\usepackage{amsmath}
\usepackage{pictexwd,dcpic}

\begin{document}

\title{Non-associative slave-boson decomposition}
\author{Vladimir Dzhunushaliev
\footnote{Senior Associate of the Abdus Salam ICTP}} 
\email{dzhun@krsu.edu.kg} \affiliation{Dept. Phys. and Microel. 
Engineer., Kyrgyz-Russian Slavic University, Bishkek, Kievskaya Str. 
44, 720021, Kyrgyz Republic}


\begin{abstract}
The operator constraint 
$f_{i\uparrow}^\dagger f_{i\uparrow} + f_{i\downarrow}^\dagger f_{i\downarrow} + 
b^\dagger_{i} b_{i}= 1$ in $t$-$J$ model of High-T$_c$ superconductivity is considered. It is shown that the constraint can be resolved by introdusing a non-associative operator. In this case the constraint is an antiassociative generating relation of a new algebra. Similar constraint is offered for splitting the gluon. 
\end{abstract}

\maketitle

\section{Introduction}

Everybody knows that the algebra of non-perturbative operators in quantum theory exists but nobody knows its exact form. In this paper the idea is discussed that the constraint \eqref{1-40} in $t$-$J$ model is a new generating relation for an algebra of operators the product of which gives us the electron operator. On the perturbative level the algebra of quantum fields is defined by canonical (anti)commutative relationships. The algebra of non-perturbative operators should be more complicated and should be generated not only by canonical (annti)commutative relationships but should exist other generating relationships as well. In this paper we discuss the idea that the constraint \eqref{1-40} is an antiassociator in a non-associative algebra of quantum non-perturbative operators. 

\section{Operator properties of $t$-$J$ model}

It is widely believed that the low energy physics of High-T$_c$ cuprates (for review, see Ref.
\cite{lee}) is described in terms of 
$t$-$J$ type model, which is given by \cite{LN9221}
\begin{equation}
	H = \sum \limits_{i,j} J\left(
	{{S}}_{i}\cdot {{S}}_{j}-\frac{1}{4} n_{i} n_{j} \right)
	-\sum_{i,j} t_{ij}
	\left(c_{i\sigma}^\dagger
	c_{j\sigma}+{\rm H.c.}\right)
\label{1-10}
\end{equation}
where $t_{ij}=t$, $t'$, $t''$ for the nearest, second nearest and 3rd nearest
neighbor pairs, respectively. The effect of the strong Coulomb repulsion is
represented by the fact that the electron operators $c^\dagger_{i\sigma}$ and
$c_{i\sigma}$ are the projected ones, where the double occupation is
forbidden.  This is written as the inequality
\begin{equation}
	\sum_{\sigma} c^\dagger_{i\sigma} c_{i \sigma} \le 1 
\label{1-20}
\end{equation}
which is very difficult to handle.  A powerful method to treat this constraint
is so called the slave-boson method \cite{B7675,C8435}.  In this approach the electron operator is
represented as
\begin{equation}
	c^\dagger_{i\sigma} = f_{i\sigma}^\dagger b_{i} 
\label{1-30}
\end{equation}
where $f_{i\sigma}^\dagger$, $f_{i \sigma}$ are the fermion operators, while $b_{i}$ is the
slave-boson operator.  This representation together with the constraint 
\begin{equation}
	f_{i\uparrow}^\dagger f_{i\uparrow} + f_{i\downarrow}^\dagger f_{i\downarrow} + 
	b^\dagger_{i} b_{i} = 1
\label{1-40}
\end{equation}
reproduces all the algebra of the electron operators. The physical meaning of the operators $f$ and
$b$ is unclear: do exist these fields or not ? 
\par 
In this paper we would like to show that the constraint \eqref{1-40} can be considered as a
generating relationship for a new algebra of operators from which the electron operator is
constructed. 

\section{A simple physical consideration}

At this stage we ignore all indixes in the constraint \eqref{1-40}. In this case Eq. \eqref{1-40}
has the form 
\begin{equation}
	f^\dagger f + b^\dagger b = 1.
\label{2-10}
\end{equation}
Now we want to compare this equation with one of generating relations in a non-associative algebra proposed in \cite{dzhun1, dzhun2}. Let us assume that there exists a non-associative algebra $G$. The algebra $G$ is generated with associators and antiassociators (shortly speaking $(\pm)$associators). One of the $(-)$associators has the form 
\begin{equation}
	\left( Q_1 Q_2 \right) Q_3 + Q_1 \left( Q_2 Q_3 \right) = \text{(something)}
\label{2-20}
\end{equation}
where $Q_i, i=1,2,3$ are non-associative operators and it is the antiassociator \eqref{sec3-40} from
Section \ref{three_op}. In the simplest case $\text{(something)} = 1$.
Let us
compare both Eq's \eqref{1-40} and \eqref{2-20}. It is easy to see that they are \emph{identical}
if we assign 
\begin{equation}
\begin{split}
	f^\dagger =& Q_1 Q_2, \quad f = Q_3 ,
\\
	b^\dagger =& Q_1, \quad b = Q_2 Q_3.
\end{split}
\label{2-30}
\end{equation}
Immediately we see that using Eq's \eqref{2-30}, Eq. \eqref{2-20} can be rewritten in the following
way
\begin{equation}
	\left( b^\dagger Q_2 \right) f + b^\dagger \left( Q_2 f \right) = 1
\label{2-35}
\end{equation}
here for the simplicity we assume that r.h.s of \eqref{2-20} is unity. Eq. \eqref{2-35} tells us
that the constraint \eqref{2-10} can be resolved in a non-associative algebra by the introduction of
a non-associative operator $Q_2$. 
\par 
Thus the idea presented here is that the slave-boson decomposition is nothing else than 
\emph{the decomposition of an associative operator on non-associative operators.} Such
non-associative algebra should have an associative subalgebra with the elements $c$ given as 
\begin{equation}
	c = f^\dagger b
\label{2-36}
\end{equation}
where $f$ and $b$ are non-associative operators. In other words in a non-associative algebra $G$ the observables $c$ have the slave-boson decomposition \eqref{2-36} where  non-associative operators $f$ and $b$ are unobservables quantities. 
\par 
Now one can restore the spin index $\sigma$ and write 
\begin{equation}
\begin{split}
	f^\dagger_{\sigma} =& Q_1 Q_{2 \sigma}, \quad f_{\sigma} = Q_{3 \sigma} ,
\\
	b^\dagger =& Q_1, \quad b = \sum_\sigma Q_{2 \sigma} Q_{3 \sigma}
\end{split}
\label{2-40}
\end{equation}
where $\sigma = \left \{ \uparrow, \downarrow \right \}$ is the spin index. In this case the $(-)$associator \eqref{2-20} has the form 
\begin{equation}
	\sum_\sigma \left( Q_1 Q_{2 \sigma} \right) Q_{3 \sigma} + 
	Q_1 \left( \sum_\sigma Q_{2 \sigma} Q_{3 \sigma} \right) = \text{(something)}. 
\label{2-45}
\end{equation}
Analogously to Eq. \eqref{2-35} the non-associative solution of \eqref{1-40} is 
\begin{equation}
	\sum_\sigma \left( b^\dagger Q_{2 \sigma} \right) f_{\sigma} + 
	b^\dagger \left( \sum_\sigma Q_{2 \sigma} f_{\sigma} \right) = \text{(something)}
\label{2-47}
\end{equation}
where $Q_{2 \sigma}$ is unknown non-associative operator. 
\par 
The problem of such interpretation of the constraint \eqref{1-40} is evidently: there exists a
non-associative algebra with $(-)$associators \eqref{2-20}, \eqref{2-45} ? 

\section{Splitting the gluon ?}

The title of this section is the same as the title of Ref. \cite{Niemi:2005qs} where it is shown
that there exists the decomposition of gluon in Yang-Mills gauge theory similar to the
slave-boson decomposition in High-T$_c$ superconductivity. In Ref. \cite{Chernodub} the similar
construction (spin-charge separation) for the gauge boson is offered as well. The physical ground is
that in both cases we are dealing with strong interactions: between electrons in High-T$_c$
superconductivity and gauge bosons in quantum chromodynamics. 
\par
In Ref. \cite{Niemi:2005qs} the slave-boson decomposition of the $SU(2)$ gauge field $A^a_\mu$
($a=1,2,3$ and $\mu=0,1,2,3$) proceeds as follows \cite{ludvig1,oma1}: at first the 
diagonal Cartan component $A^3_\mu = A_\mu$ from the off-diagonal components $A^{1,2}_\mu$ is
separated, and combined the latter into the complex field $W_\mu = A^1_\mu + i A^2_\mu$. Then
a complex vector field
$\vec{ e}_\mu$ with
\begin{equation}
	\vec{ e}_\mu \vec{ e}_\mu = 0 \qquad \text{and}
	\qquad \vec{ e}_\mu \vec{ e}^*_\mu = 1 
\label{4-01}
\end{equation} 
is introduced; two spinless complex scalar fields $\psi_1$ and $\psi_2$ are introduced as well. 
The ensuing decomposition of $W_\mu$ is \cite{ludvig1}
\begin{equation}
W_\mu = A^1_\mu + i A^2_\mu = \psi_1 \vec{ e}_\mu + 
\psi_2^* \vec{ e}^*_\mu.
\label{4-02}
\end{equation} 
This is a direct analogue of Eq.~(\ref{1-40}), a decomposition of $W_\mu$ into spinless
bosonic scalars $\psi_{1,2}$ which describe the gluonic holons that carry the color charge of the
$W_\mu$, and a color-neutral spin-one vector $\vec{ e}_\mu$ which is the gluonic spinon that carries
the statistical spin degrees of freedom of $W_\mu$. 
\par 
In Ref. \cite{Chernodub} the spin-charge separation of SU(2) gauge potential is given by a 
little another way 
\begin{equation}
	A^a_\mu = e^i_\mu \Phi^{ia}
\label{4-10}
\end{equation}
where $a=1,2,3$ is the SU(2) color index; $i=1,2,3$ is an inner index and $\mu = 0,1,2,3$ is the Lorentzian index. The decompositions \eqref{4-02} and \eqref{4-10} are the decompositions of the classical fields. If we trust the quantum slave-boson decomposition for strongly interacting electrons in High-T$_c$ superconductivity then we can apply this idea for the strongly interacting SU(3) gauge potential. In this case the non-perturbative operator $\hat A^B_\mu$ can be decomposed in the following way 
\begin{equation}
	\hat A^B_\mu = \hat e^i_\mu \hat \Phi^{iB}
\label{4-20}
\end{equation}
here we follow to the decomposition \eqref{4-10}; $B=1, \cdots , 8$ is the SU(3) color index. Following to the slave-boson idea we assume that there is a constraint 
\begin{equation}
	\sum_{i,\mu} \hat e^{\dagger i}_\mu \hat e^{i\mu} + 
	\sum_{i,B} \hat \Phi^{\dagger iB} \hat \Phi^{iB} = 1 .
\label{4-30}
\end{equation}
Analogously to \eqref{1-40} the resolution of this constraint is a $(-)$ associator 
\begin{equation}
	\sum_{i, j, \mu , B} \left( \hat \Phi^{\dagger jB} \hat Q^{jiB}_\mu \right) 
	\hat e^{i \mu} + \sum_{i, j, \mu , B} \hat \Phi^{\dagger iB} 
	\left( \hat Q^{ijB}_\mu \hat 	e^{j \mu} \right) = 1
\label{4-40}
\end{equation}
with 
\begin{equation}
	\hat e^{\dagger i}_\mu= \sum_{j, B} 
	\left( \hat \Phi^{\dagger j B} \hat Q^{jiB}_\mu \right),
	\quad 
	\hat \Phi^{i B} = \sum_{j, \mu} 
	\left( \hat Q^{ijB}_\mu \hat e^{j \mu} \right)
\label{4-50}
\end{equation}
where $\hat Q^{ijB}_\mu$ is an unknown non-associative operator. 
\par 
In the next section we will present a non-associative algebra where $(\pm)$associators are given on the level of the product of three operators. 

\section{A non-associative algebra}

In this section we will follow to Ref. \cite{dzhun1}. At first we would like to note that:
(a) non-associative algebras proposed in \cite{dzhun1, dzhun2} are the operator generalization of the octonions; (b) the full definition of this algebra is jet unknown:
here we give the $(\pm)$associators for the product of three operators only. 

\subsection{Octonions}

In this subsection we give a very short description what is it the octonion numbers. 
Let split-octonion numbers are designeted as $\tilde{q}_i$ and $\tilde{Q}_j$. In Table \ref{oct} we
present 
the multiplication rule of the split-octonions numbers $\tilde{q}_i, \tilde{Q}_j$  and $I$. 
\begin{table}[h]
\begin{tabular}{|c|c|c|c|c|c|c|c|c|}                                                
\hline
&$\tilde{q}_1$ & $\tilde{q}_2$ & $\tilde{q}_3$ & $\tilde{Q}_1$  & $\tilde{Q}_2$ & $\tilde{Q}_3$ & $I$         \\ 
\hline 
$\tilde{q}_1$& $-1  $ & $\tilde{q}_3$  & $-\tilde{q}_2$  & $-I$ & $\tilde{Q}_3$  & $-\tilde{Q}_2$ & $\tilde{Q}_1$ \\ 
\hline
$\tilde{q}_2$ & $-\tilde{q}_3$ & $-1$   & $\tilde{q}_1$  & $-\tilde{Q}_3$  & $-I$ & $\tilde{Q}_1$  & $\tilde{Q}_2$    \\ 
\hline
$\tilde{q}_3$ & $\tilde{q}_2$ & $-\tilde{q}_1$ & $-1$   & $\tilde{Q}_2$  & $-\tilde{Q}_1$  & $-I$ & $\tilde{Q}_3$    \\ 
\hline
$\tilde{Q}_1$ & $I $ & $\tilde{Q}_3$ & $-\tilde{Q}_2$ & $ 1$   & $-\tilde{q}_3$  & $\tilde{q}_2$  & $\tilde{q}_1$    \\ 
\hline
$\tilde{Q}_2$ & $-\tilde{Q}_3$ & $I$  & $\tilde{Q}_1$ & $\tilde{q}_3$ & $ 1$   & $-\tilde{q}_1$  & $\tilde{q}_2$    \\ 
\hline
$\tilde{Q}_3$ & $\tilde{Q}_2 $ & $-\tilde{Q}_1$ & $I$  & $-\tilde{q}_2$ & $\tilde{q}_1$ & $ 1$   & $\tilde{q}_3$    \\ 
\hline
$I$ & $-\tilde{Q}_1 $ & $-\tilde{Q}_2$  & $-\tilde{Q}_3$ & $-\tilde{q}_1$  & $-\tilde{q}_2$ & $-\tilde{q}_3$ & $ 1$    \\ 
\hline
\end{tabular}
\caption{The split-octonions multiplication table.} 
\label{oct}
\end{table}

\subsection{Quantum ($\pm$)associators for the product of three operators}
\label{three_op}

In this section we present the quantum ($\pm$)associators for the product of three operators. The anticommutators are
\begin{eqnarray}
    \left\{ q_i q_j \right\}_+ &=& 0,
\label{sec2-72}\\
    \left\{ q_i Q_j \right\}_+ &=& 0,
\label{sec2-74}\\
    \left\{ Q_i Q_j \right\}_+ &=& 0,
\label{sec3-10}\\
  \left\{ q_i Q_i \right\}_+ &=& 0 .
\label{sec3-20}
\end{eqnarray}
The quantum ($\pm$)associators with different indices $m \neq n, n \neq p, p \neq m $ are
\begin{eqnarray}
    \left\{ q_m,q_n,q_p \right\}_- &=& 
    \left( q_m q_n \right) q_p - q_m\left( q_n q_p \right) = 0 ,
\label{sec3-30}\\
    \left\{ Q_m,Q_n,Q_p \right\}_+ &=& 
    \left( Q_mQ_n \right) Q_p + Q_m\left( Q_n Q_p \right) =
    \epsilon_{mnp} \mathcal{H}_{3,1} ,
\label{sec3-40}\\
    \left\{ q_m,Q_n,q_p \right\}_+ &=& 
    \left( q_m Q_n \right) q_p + q_m\left( Q_n q_p \right) =
    \epsilon_{mnp} \mathcal{H}_{3,2} ,
\label{sec3-50}\\
    \left\{ Q_m,q_n,Q_p \right\}_- &=& 
    \left( Q_m q_n \right) Q_p - Q_m\left( q_n Q_p \right) =
    \epsilon_{mnp} \mathcal{H}_{3,3} ,
\label{sec3-60}\\
    \left\{ q_m,q_n,Q_p \right\}_- &=& 
    \left( q_m q_n \right) Q_p - q_m\left( q_n Q_p \right) =
    \epsilon_{mnp} \mathcal{H}_{3,4} ,
\label{sec3-70}\\
    \left\{ Q_m,q_n,q_p \right\}_- &=& 
    \left( Q_m q_n \right) q_p - Q_m\left( q_n q_p \right) =
    \epsilon_{mnp} \mathcal{H}_{3,5} ,
\label{sec3-80}\\
    \left\{ q_m,Q_n,Q_p \right\}_- &=& 
    \left( q_m Q_n \right) Q_p - q_m\left( Q_n Q_p \right) =
    \epsilon_{mnp} \mathcal{H}_{3,6} ,
\label{sec3-90}\\
    \left\{ Q_m,Q_n,q_p \right\}_- &=& 
    \left( Q_m Q_n \right) q_p - Q_m\left( Q_n q_p \right) =
    \epsilon_{mnp} \mathcal{H}_{3,7}
\label{sec3-100}
\end{eqnarray}
where $\mathcal{H}_{3,i}$ are operators (but may be they are numbers); Eq. \eqref{sec3-30} means that the quaternionic - like subalgebra spanned on $q_1, q_2, q_3$ is the associative algebra. The quantum antiassociators for the product of three operators, such as $q(Qq)$ or $Q(qQ)$, having 
two different indices $m \neq n$ are
\begin{eqnarray}
    \left\{ q_m,Q_n,q_n \right\}_+ &=& 
    q_m\left( Q_n q_n \right) + \left( q_m Q_n \right) q_n = \mathcal{H}_{3,8}(m,n) ,
\label{sec3-120}\\
    \left\{ q_n,Q_n,q_m \right\}_+ &=& 
    q_n\left( Q_n q_m \right) + \left( q_n Q_n \right) q_m = \mathcal{H}_{3,9}(m,n) ,
\label{sec3-130}\\
    \left\{ Q_m,q_n,Q_n \right\}_+ &=& 
    Q_m\left( q_n Q_n \right) + \left( Q_m q_n \right) Q_n = \mathcal{H}_{3,10}(m,n) ,
\label{sec3-140}\\
    \left\{ Q_n,q_n,Q_m \right\}_+ &=& 
    Q_n\left( q_n Q_m \right) + \left( Q_n q_n \right) Q_m = \mathcal{H}_{3,11}(m,n) .
\label{sec3-150}
\end{eqnarray}
The quantum associators, such as $q(QQ)$ or $Q(qq)$, and with two different indices $m \neq n$ are
\begin{eqnarray}
    \left\{ q_m, Q_m, Q_n \right\}_- &=& 
    \left( q_m Q_m \right) Q_n - q_m\left( Q_m Q_n \right) = \mathcal{H}_{3,12}(m,n) ,
\label{sec3-160}\\
    \left\{ q_m, Q_n, Q_m \right\}_- &=& 
   \left( q_m Q_n \right) Q_m - q_m\left( Q_n Q_m \right) = \mathcal{H}_{3,13}(m,n) ,
\label{sec3-170}\\
    \left\{ Q_n, Q_m, q_m \right\}_- &=& 
    \left( Q_n Q_m \right) q_m - Q_n\left( Q_m q_m \right) = \mathcal{H}_{3,14}(m,n) ,
\label{sec3-180}\\
    \left\{ Q_m, Q_n, q_m \right\}_- &=& 
    \left( Q_m Q_n \right) q_m - Q_m\left( Q_n q_m \right) = \mathcal{H}_{3,15}(m,n) ,
\label{sec3-190}\\
    \left\{ Q_m, q_m, q_n \right\}_- &=& 
    \left( Q_m q_m \right) q_n - Q_m\left( q_m q_n \right) = \mathcal{H}_{3,16}(m,n) ,
\label{sec3-200}\\
    \left\{ Q_m, q_n, q_m \right\}_- &=& 
    \left( Q_m q_n \right) q_m - Q_m\left( q_n q_m \right) = \mathcal{H}_{3,17}(m,n) ,
\label{sec3-210}\\
    \left\{ q_n, q_m, Q_m \right\}_- &=& 
    \left( q_n q_m \right) Q_m - q_n\left( q_m Q_m \right) = \mathcal{H}_{3,18}(m,n) ,
\label{sec3-220}\\
    \left\{ q_m, q_n, Q_m \right\}_- &=& 
    \left( q_m q_n \right) Q_m - q_m\left( q_n Q_m \right) = \mathcal{H}_{3,19}(m,n) 
\label{sec3-230}
\end{eqnarray}
where $\mathcal{H}_{3,i}(m,n)$ are operators (but may be they are numbers). The alternativity properties are
\begin{eqnarray}
    \left\{ q_n, q_n, Q_m \right\}_- &=& 
    \left( q_n q_n \right) Q_m - q_n \left( q_n Q_m \right) = 0,
\label{sec3-240}\\
    \left\{ q_n, Q_m, q_n \right\}_- &=& 
    \left( q_n Q_m \right) q_n - q_n \left( Q_m q_n \right) = 0,
\label{sec3-250}\\
    \left\{ Q_m, q_n, q_n \right\}_- &=& 
    \left( Q_m q_n \right) q_n - Q_m \left( q_n q_n \right) = 0.
\label{sec3-260}
\end{eqnarray}

\subsection{Self - consistency of quantum ($\pm$)associators for the product of three operators}

The self-consistency of the $(\pm)$associators for the product of three operators can be proved according to the commutative diagram \eqref{sec4-10}. For this we permute the first and third factors in the product $a(bc)$ following to the commutative diagram 
\begin{equation}
\begindc{\commdiag}[3]
\obj(1,5){$a \left( bc \right)$}
\mor(4,5)(15,11){}
\mor(4,5)(15,-1){}

\obj(17,10){$a \left( cb \right)$}
\obj(32,10){$\left(ac \right) b$}
\obj(47,10){$\left(ca \right) b$}
\obj(62,10){$c \left(ab \right)$}
\mor(19,10)(30,10){}
\mor(34,10)(45,10){}
\mor(49,10)(60,10){}

\obj(17,0){$\left( ab \right) c$}
\obj(32,0){$\left( ba \right) c$}
\obj(47,0){$b \left( ac \right)$}
\obj(62,0){$b \left(ca \right)$}
\obj(77,0){$\left( bc \right) a$}
\obj(92,0){$\left( cb \right) a$}
\mor(19,0)(30,0){}
\mor(34,0)(45,0){}
\mor(49,0)(60,0){}
\mor(64,0)(75,0){}
\mor(79,0)(90,0){}

\obj(110,5){$c \left( ba \right)$}
\mor(95,0)(107,5){}
\mor(65,11)(107,5){}
\enddc
\label{sec4-10}
\end{equation}
In Ref's.~\cite{dzhun1,dzhun2} it is shown that the $(\pm)$associators with the product of three operators are self-consistent. On this level it is impossible to define operators $\mathcal H_{3,i}$.

\section{Outlook}

In this paper we have shown that the constraint \eqref{1-40} can be resolved by an  unexpected manner: the constraint is a generating relation for a new non-associative  algebra. The bilinear combination of operators of this algebra make up an operator of an observable physical quantity. Probably: (a) such decomposition can be done only in the case if these physical quantities have strong interaction; (b) there exists an associative subalgebra of the above-mentioned non-associative algebra where associative operators can be decomposed by non-associative operators following to the slave-boson way. In Ref. \cite{dzhun2} such observable physical quantities are called white (colorless) operators. One can say that this hypotesized decomposition of an associative operator on non-associative operators (similar Eq's \eqref{1-30} and \eqref{4-20}) is in some sense the generalization of quark confinement hypothesis: in both cases we have observable physical quantities which are build from unobservable quantities. 

\section{Acknowledgments}

The author thanks D. Ebert, M. Mueller-Preussker and the other colleagues of the Particle Theory
Group of the Humboldt University for kind hospitality. This work has been supported by DAAD.

\end{document}